\documentclass[12pt]{article}
\usepackage{amsmath}
\usepackage{amssymb}
\usepackage{mathtext}
\usepackage{graphicx}
\begin{document}
\begin{centering}
\section*{SUPERINTEGRABLE SYSTEMS ON SPHERE}
\begin{centering}
A.\,V.\,BORISOV, I.\,S.\,MAMAEV\\
Institute of Computer Science,\\
Udmurt State University,\\
Universitetskaya 1, 426034, Izhevsk, Russia\\
E-mail: borisov@rcd.ru, mamaev@rcd.ru\\
\end{centering}\bigskip
\end{centering}

\begin{abstract}
We consider various generalizations of the Kepler problem to
three-dimensional sphere~$S^3$, a compact space of constant curvature.
These generalizations include, among other things, addition of a spherical
analog of the magnetic monopole (the Poincar\'e--Appell system) and
addition of a more complicated field, which itself is a generalization of
the MICZ-system. The mentioned systems are integrable~--- in fact,
superintegrable. The latter is due to the vector integral, which is
analogous to the Laplace--Runge--Lenz vector. We offer a classification of
the motions and consider a trajectory isomorphism between planar and
spatial motions. The presented results can be easily extended to
Lobachevsky space~$L^3$.
\end{abstract}

\section{The Kepler problem in $\mathbb{R}^3$}

Consider the Kepler problem: a mass point (of unit mass, without losing in
generality) moves in the Newtonian field of a fixed center; the intensity
of the gravitational interaction~$\gamma$ is constant.

In this problem, in addition to the integral of energy
\begin{equation}
\label{e12-2}
 H_0^{}
= \frac{1}{2} \sum_{i=1}^3 \dot{q}_i^2 + U
\end{equation}
 and the vector integral of
angular momentum
\begin{equation}
\label{e12-3}
\boldsymbol M = \boldsymbol q \times \boldsymbol {\dot q},
\end{equation}
 the equations in
three-dimensional Euclidean space~$\mathbb{R}^3 = \{q_1^{}
,\,q_2^{},\,q_3^{}\}$, \vskip0.5mm\noindent
\begin{equation}
\label{e12-1}
\ddot q_i^{} =
\frac{\partial{U}}{\partial {q_i}},\quad U = -\frac{\gamma}{r},\quad r^2 = q_1^2 + q_2^2 +
q_3^2,\quad \gamma= const,
\end{equation}
\noindent have one more remarkable
vector integral, which is due to certain hidden symmetry of the Kepler
problem. This vector integral is called the Laplace--Runge--Lenz
vector~$\boldsymbol A=(A_1^{},\,A_2^{},\,A_3^{})$. It exists only in the case of
Newtonian potential (of all the central potentials) and can be written as
follows:
\begin{equation}
\label{e12-4}
\boldsymbol A = \boldsymbol M \times \dot{\boldsymbol q} + \frac{\gamma}{r}\boldsymbol q.
\end{equation}
Introducing the momenta~$\boldsymbol p = \dot{\boldsymbol q}$, we can rewrite equations
\eqref{e12-1} and integrals \eqref{e12-2}, \eqref{e12-3},
and~\eqref{e12-4} in the canonical form:
\begin{equation}
\label{e12-5}
 \dot{\boldsymbol p} = \frac{\partial {H_0}}{\partial {\boldsymbol q}},\quad
\dot{\boldsymbol q} = - \frac{\partial{H_0}}{\partial{\boldsymbol p}},\quad \boldsymbol p,\,\boldsymbol q
\in \mathbb{R}^3.
\end{equation}

The Poisson brackets for the components of the integrals~$\boldsymbol M$ and~$\boldsymbol A$
are
\begin{equation}
\label{e12-6}
 \{M_i, M_j\} = \varepsilon_{ijk} M_k,\quad
\{M_i, A_j\} = \varepsilon_{ijk} A_k,\quad \{A_i,\,A_j\}
= -2h\varepsilon_{ijk} M_k,
\end{equation}
 where $h$ is the constant of
energy~\eqref{e12-2}, $h=\frac{1}{2}\boldsymbol p^2 - \frac{\gamma}{r}$, and
$\varepsilon_{ijk}^{}$ is the Levi--Civita symbol. Depending on the value of~$h$,
the algebra of integrals~\eqref{e12-6} is either~$so(4)$ (when~$h<0$)
or~$so(3,\,1)$ (when~$h>0$).

Note that, since $(\boldsymbol M,\,\boldsymbol A) = 0$, $\boldsymbol A$ is
always in the plane of the trajectory. Besides, the vector's direction
coincides with the direction of the ellipse's major axis, while its
absolute value is proportional to the eccentricity.

The algebra of integrals~\eqref{e12-6} is an algebra, under which the
Kepler problem is invariant. Invariance under a global group of
transformations (i.\,e., for example, under group~$SO(4)$ for~${h<0}$) was
studied by V.\,A.\,Fok~\cite{b12-1}, G.\,Gy\"{o}rgyi~\cite{b12-2} and
J.\,Moser~\cite{b12-3}. The latter work contains the most general result,
which shows that even in the~$n$-dimensional case, the constant energy
surface (for~$h<0$) after suitable regularization is topologically
equivalent to the bundle of unit vectors tangent to~$n$-dimensional
sphere~$S^n$.

Note also that the principal dynamical effect of a redundant algebra of
integrals~\eqref{e12-6} is the fact that the trajectories of
system~\eqref{e12-1} are closed in the configurational and phase spaces.

\section{The MICZ-system in $\mathbb{R}^3$. Appell's problem}

Consider one more generalization of the Kepler problem, for which an
analog of integral~\eqref{e12-4} exists. To this end, in the phase space
$T^\ast \mathbb{R}^3$ we specify a noncanonical Poisson bracket
\begin{equation}
\label{e12-7}
\{q_i^{},\,q_j^{}\} = 0,\quad \{q_i^{},\,p_j^{}\} =
\delta_{ij}^{},\quad \{p_i^{},\,p_k^{}\} = - \mu \varepsilon_{ijk}^{}
\frac{q_k^{}}{r^3}
\end{equation}
and a Hamiltonian
\begin{equation}
\label{e12-8}
H_1^{} =\frac12
\sum_{i=1}^3 p_i^2 - \frac{\gamma}{r} + \frac{\mu^2}{2r^2}, \quad
\gamma,\,\mu = const.
\end{equation}

{\bf Remark.} {\it
This system (the differential equations of motion) can as well be obtained
with the standard canonical bracket, but in this case the Hamiltonian
would contain terms linear in momenta.}

Equations \eqref{e12-7}, \eqref{e12-8} define the MICZ-system (McIntosh-
Cis\-ne\-ros-Zwanziger); it describes a particle's motion in the
asymptotic field of a self-dual monopole~\cite{b12-4}. It was formally
studied by Zwanziger~\cite{b12-5}, McIntosh and Cis\-ne\-ros~\cite{b12-6}
without any relevant physical interpretation (see also~\cite{b12-19}).

Consider some special cases of~\eqref{e12-7}, \eqref{e12-8}. The Kepler
problem can be obtained if we put~$\mu=0$. Putting~$\gamma=0$ and $\mu=0$
in the Hamiltonian~\eqref{e12-8}, not in the bracket~\eqref{e12-7}, we
have the classical integrable Poincar\'e problem of a particle moving in
the field of a magnetic monopole. As it was shown by Poincar\'e, the
particle's trajectories in this case are geodesics of a circular cone.

\enlargethispage*{\baselineskip}

P.\,Appell considered a more general problem of a particle moving in the
field of a Newtonian center and in the field of a magnetic monopole,
assuming that the center and the monopole coincide~\cite{b12-7}. This
problem occurs if~$\mu=0$ in~\eqref{e12-8} (in the bracket~\eqref{e12-7},
however,~$\mu\ne 0$). In this case, the trajectory is a conic section in
the involute of the circular cone, while the integral of areas is
preserved during the motion. On the cone itself, the trajectories are,
generally, not closed.

There is no analog of the integral~$\boldsymbol A$ \eqref{e12-4} in the Poincar\'e
and Appell problems, but it exists for the system \eqref{e12-7},
\eqref{e12-8}. As for the vector integral of angular momentum~$\boldsymbol M$, it
exists for all the above problems. Indeed~\cite{b12-8}, the vector
functions
\begin{equation}
\label{e12-9}
\boldsymbol M=\boldsymbol q\times\boldsymbol p + \mu\frac{\boldsymbol q}{r},
\end{equation}

\begin{equation}
\label{e12-10} \boldsymbol A=\frac{1}{\sqrt{|2H_1^{}|}}\left(\boldsymbol
p\times\boldsymbol M-\frac{\boldsymbol q}{r}\right)
\end{equation}
form the algebra of integrals of~\eqref{e12-7}, \eqref{e12-8}, which is
isomorphic to~$so(4)$ for~${H_1^{} < 0}$ and to~$so(3,1)$ for~$H_1^{}> 0$.

Again the trajectories are conic sections, and, since~$(\boldsymbol M,\,\boldsymbol q/r) = -
\mu$, belong to the circular cone with cone angle~$\theta = \arccos
\mu/|\boldsymbol M|$ and the axis of symmetry, defined by~$\boldsymbol M$.

Various generalizations of the Laplace--Runge--Lenz integrals to
dynamical systems in the Euclidean space were studied in~\cite{b12-15}.

\section[The Kepler problem on three-dimensional sphere~$S^3$ (and on
Lobachevsky space~$L^3$)]{The Kepler problem on three-dimensional
sphere~$S^3$ (and on Lobachevsky space~$L^3$) \markright{\thesection. The
Kepler problem on three-dimensional sphere~$S^3$}} \markright{\thesection.
The Kepler problem on three-dimensional sphere~$S^3$}

Consider analogs of the Kepler problem in some simple non-Euclidean spaces
of constant curvature~--- three-dimensional sphere~$S^3$ and Lobachevsky
space. We will discuss the spherical case in more detail. However, all the
results, after appropriate revision, can be extended to Lobachevsky space.

Let the three-dimensional sphere~$S^3$ be embedded into the
four-dimensional Euclidean space~$\mathbb{R}^4 =
\{q_0^{},\,q_1^{},\,q_2^{},\,q_3^{}\}$ and given by equation~
\begin{equation}
\label{e12-11}
q_0^2 + q_1^2 + q_2^2 + q_3^2 = R^2,
\end{equation}
where $R$ is the sphere's radius.

Introduce spherical coordinates on $S^3$:
\begin{equation}
\label{e12-29}
\begin{gathered}
q_0^{}  = R\cos \theta,\quad q_1^{} = R \sin \theta \cos \varphi, \\
q_2^{}  = R\sin \theta \sin \varphi \cos \psi, \quad
q_3^{}  = R\sin \theta \sin \varphi \sin \psi.
\end{gathered}
\end{equation}

Consider the motion of a particle in the field of a Newtonian center,
placed at one of the poles,~$\theta = 0$, of the three-dimensional sphere. It
is well known~\cite{b12-10,b12-11,b12-12,b12-13,b12-16,b12-17} that the
analog of Newtonian potential on sphere is~
\begin{equation}
\label{e12-12}
U = - \gamma \cot
\theta = - \gamma \frac{q_0^{}}{|\boldsymbol q|}, \quad \boldsymbol q^2 = q_1^2+q_2^2+q_3^2,\quad \boldsymbol q
= (q_1^{},\,q_2^{},\,q_3^{}).
\end{equation}

Recall that the potential~\eqref{e12-12} can be obtained either by solving
the Laplace--Beltrami equation on sphere~$S^3$ (see below,~\eqref{Bor}).
This equation is invariant under group~$SO(3)$ and has a singularity at
the pole~$\theta = 0$, or by extending Bertrand's theorem to
sphere~\cite{b12-9,b12-10}.

In independent coordinates $\boldsymbol q = (q_1, q_2, q_3)$, the
Lagrangian of the problem in question is
\begin{equation}
\label{e12-13}
L = \frac{1}{2}{(\boldsymbol {\dot q}^2 + q_0^{-2}{(\boldsymbol q, \boldsymbol {\dot q})}^2)}^2 - U(\boldsymbol q),
\end{equation} where $q_0^{}$ is
found from~\eqref{e12-11}, namely,~$q_0^{} = \pm \sqrt{R^2 - \boldsymbol q^2}$.
After introducing the momenta~
\begin{equation}
\label{e12-14}
\boldsymbol p = \frac{\partial L}{\partial{\boldsymbol {\dot q}}} =
\boldsymbol {\dot q} + \frac{(\boldsymbol q, \boldsymbol {\dot q})}{\sqrt{R^2 - \boldsymbol q^2}}\boldsymbol q,
\end{equation}
the equations of
motion can be represented in the canonical Hamiltonian form with
Hamiltonian
\begin{equation}
\label{e12-15}
H = \frac{1}{2} \boldsymbol p^2 - \frac{1}{2R^2}(\boldsymbol p,\,\boldsymbol q)^2 +
V(\boldsymbol q).
\end{equation}
These equations have a vector integral of angular momentum
\begin{equation}
\label{e12-16}
\boldsymbol M = \boldsymbol p \times \boldsymbol q = \boldsymbol {\dot q} \times \boldsymbol q
\end{equation}
(it exists as well for
every ``central'' potential~$V$ that depends only on~$|\boldsymbol q| = \sqrt{q_1^2
+ q_2^{} + q_3^2}$) and an analog of the Laplace--Runge--Lenz integral
\begin{equation}
\label{12-17}
\boldsymbol A = q_0^{} \boldsymbol p \times \boldsymbol M + \gamma R^2 \frac{\boldsymbol q}{|\boldsymbol q|}.
\end{equation}

The components of $M_i^{}$ and $R_i^{}$ commute in the following way:
\begin{equation}
\label{e12-18}
\begin{gathered}
\{M_i^{},\,M_j^{}\} = - \varepsilon_{ijk}^{} M_k^{},\quad \{M_i^{},\,A_j^{}\} = -
\varepsilon_{ijk}^{} A_k^{},\\
\{A_i^{}\,\,A_j^{}\} = 2 (R^2 h - \boldsymbol M^2)\varepsilon_{ijk}^{} M_k^{}.
\end{gathered}
\end{equation}
(This algebra was discussed in several papers~\cite{b12-11,b12-12}.)

The Casimir functions of the nonlinear Poisson structure~\eqref{e12-18}
are
\begin{equation}
\label{e12-19}
F_1^{} = (\boldsymbol M,\boldsymbol A),\quad F_2^{} = \boldsymbol A^2 - \frac{2h}\lambda\boldsymbol M^2 +
(\boldsymbol M^2)^2,
\end{equation}
while its symplectic leaf is four-dimensional (i.\,e., the
rank of~\eqref{e12-18} is four). Here,~$\lambda=1/{R^2}$ is the curvature of
the space. For real motions, the Kepler problem gives
\begin{equation}
\label{e12-20_a}
F_1^{} =0,\quad F_2^{} = \gamma^2 R^4.
\end{equation}
The compactness of the symplectic
leaf \eqref{e12-20_a} is defined by the curvature of space,~$\lambda$, and the
value of the constant of energy:
\begin{enumerate}

\item[1.] When $\lambda = 0$: $h < 0$~--- compact,~$h \ge 0$~--- noncompact.

\item[2.] When $\lambda > 0$: always compact.

\item[3.] When $\lambda < 0$: $h < 0$ and $h^2 > \gamma^2$~--- the
leaf~\eqref{e12-20_a} is disconnected, one component is compact, while the
other is noncompact; $h > - \gamma$~--- the leaf is connected, but
noncompact.
\end{enumerate}

The trajectories of the Kepler problem on sphere (and pseudosphere) are
conic sections, the generalization of Kepler's laws to this case was done
in~\cite{b12-9,b12-13,b12-14}. In the paper~\cite{b12-20}, bifurcational
analysis of the Kepler problem on~$S^3$ and~$L^3$ was performed, and the
action-angle variables were introduced (see also~\cite{b12-21}).

\section{Generalization of the Poincar\'e and Appell problems to
$S^3$}\label{sec4}

First, we obtain the Hamiltonian form of the equations of a particle's
motion on three-dimensional sphere~$S^3$ under generalized potential
forces. Indeed, consider the Lagrangian
\begin{equation}
\label{e12-19}
L = \frac{1}{2}
\bigl(\boldsymbol {\dot q} + q_0^{-2}(\boldsymbol q, \boldsymbol {\dot q})^2\bigr) - \bigl(
\boldsymbol {\dot q}, \boldsymbol W(\boldsymbol q)\bigr) - U(\boldsymbol q),
\end{equation}
where $\boldsymbol W = \boldsymbol W(\boldsymbol q) =
(W_1^{},\,W_2^{},\,W_3^{})$ is the vector potential. Introducing the
generalized momenta
\begin{equation}
\label{e12-20}
\boldsymbol p = \frac{\partial L}{\partial \boldsymbol {\dot q}} =  \boldsymbol {\dot q} - \boldsymbol W +
\frac{(\boldsymbol q, \boldsymbol {\dot q})}{\sqrt{R^2 - \boldsymbol q^2}}\boldsymbol q,
\end{equation}
we obtain the
Hamiltonian
\begin{equation}
\label{e12-21}
 H = \frac{1}{2}(\boldsymbol p+\boldsymbol W)^2 - \frac{1}{2R^2}(\boldsymbol p +
\boldsymbol W, \boldsymbol q)^2 + U(\boldsymbol q)
\end{equation}
and the canonical Poisson bracket $(\{q_i^{},p_j^{}\} = \delta_{ij}^{})$.
Due to a number of considerations, it is more convenient to study
Hamiltonian equations written in terms of slightly modified
momenta~$\tilde{\boldsymbol p} = \boldsymbol p + \boldsymbol W$, which form
the following noncanonical Poisson brackets
\begin{equation}
\label{e12-22}
\begin{gathered}
\{\tilde{p}_i^{},\,\tilde{p}_j^{}\} = \frac{\partial {W_i}}{\partial q_j} - \frac{\partial {W_j}}{\partial q_i} =
B_{ij},\\
\{q_i^{},\,\tilde{p}_j^{}\} = \delta_{ij}, \quad \{q_i^{},\,q_j^{}\} = 0,
\end{gathered}
\end{equation}
where $\boldsymbol B = \mbox{ rot } \boldsymbol W$. The Hamiltonian~\eqref{e12-21} simplifies, in
this case, to:
\begin{equation}
\label{e12-21*}
 H= \frac{1}{2} \tilde{\boldsymbol p}^2 -
\frac{1}{R^2}(\tilde{\boldsymbol p},\,\boldsymbol q)^2 + U(\boldsymbol q).
\end{equation}

In the case of three-dimensional sphere, an analog of the vector potential
of a magnetic monopole can be obtained in the following way.

Electromagnetic field tensor in vacuum satisfies the Maxwell equations
\begin{equation}
\label{e12-23}
\begin{gathered}
\partial_\alpha F_{\beta\gamma} + \partial_\beta F_{\gamma\alpha} + \partial_\gamma
F_{\alpha\beta} = 0 \quad \alpha, \beta = 0,1,2,3\\
\frac{1}{\sqrt{-g}} \partial_\beta^{}(\sqrt{-g} F^{\alpha \beta}) = 0
\quad
\left(\partial_\alpha = \frac{\partial{}}{{\partial x_\alpha}}\right),
\end{gathered}
\end{equation}
where $\|g_{\alpha \beta}\|$ is the metric of space-time, $g =
\det\|g_{\alpha
\beta}\|$.

For $S^3$, the metric of space-time in the spherical
coordinates~\eqref{e12-29} is
\begin{equation}
\label{e12-24}
dS^2 = c^2 dt^2 - R^2\bigl(d
\theta^2 + \sin^2\theta (d\varphi^2 + \sin^2 \varphi d \psi^2)\bigr).
\end{equation}

Let $i$, $j$, $k$ stand only for the spatial indices, while~$g_*$
denotes the spatial portion of the metric, taken with the negative sign.
We will search for the solution, similar to that for a magnetic monopole
in flat space, in the form
$$
F_{0i}^{} = 0,\quad \sqrt{g_*} F^{ij} =
\varepsilon^{ijk}
\partial_k^{} f.
$$
From \eqref{e12-23} we find the equation for the
unknown function~$f$
$$
\partial_k^{}\left(\sqrt{g_*} g^{ik} \partial_i^{} f\right) = 0,$$
coinciding with the Laplace--Beltrami equation. The solution, invariant
under group $SO(3)$ (i.\,e. independent of~$\psi$, $\varphi$), satisfies
the equation
\begin{equation}
\label{Bor}
\frac{1}{\sin^2\theta} \frac{\partial}{\partial \theta}\left(\sin^2 \theta
\frac{\partial f}{\partial \theta}\right) = 0
\end{equation}
and looks as follows:
\begin{equation}
\label{e12-25}
 f = \alpha \cot\theta,\quad \alpha = const.
\end{equation}

{\bf Remark.} {\it
For Lobachevsky space $L^3$, a similar reasoning yields} $f = \alpha \coth
\theta$, $\alpha = const$.

The vector potential of the magnetic monopole $\boldsymbol W$ is found
from~$F_{ij}^{} = \partial_i^{} W_j^{} - \partial_j^{} W_i^{}$. In terms
of spherical coordinates~$(\theta,\,\varphi, \psi)$, it reads:
$$
W_\theta^{} = 0,\quad W_\varphi^{} = 0,\quad W_\psi^{} = \alpha R \cos \varphi.$$ In
terms of variables~$q_0^{}, \boldsymbol q$, it can be written as
\begin{equation}
\label{e12-26}
\boldsymbol W =
\left(0,\,\alpha \frac{q_1^{}}{|\boldsymbol q|}\frac{q_3^{}}{q_2^2 + q_3^2},\, -
\alpha
\frac{q_1^{}}{|\boldsymbol q|} \frac{q_2^{}}{q_2^2 + q_3^2}\right)\!,
\end{equation}
while
for~$\boldsymbol B = \mbox{ rot }\boldsymbol W$, we have
\begin{equation}
\label{e12-27} \boldsymbol B = -\frac{\alpha}{|\boldsymbol q|^3}\boldsymbol q.
\end{equation}

Consider a particle, moving on~$S^3$ in the field of a Newtonian center
and in the field of a magnetic monopole, the center and the monopole being
placed at the pole~$\theta = 0$. This is a spherical analog of the Appell
problem. The Hamiltonian of the problem is either \eqref{e12-21}
or~\eqref{e12-21*} with~$\boldsymbol W(\boldsymbol q)$ and~$U(\boldsymbol q)$ defined, respectively,
by~\eqref{e12-26} and~\eqref{e12-12}. Hamiltonian equations always admit
the integral of angular momentum
\begin{equation}
\label{e12-28} \boldsymbol M = \tilde{\boldsymbol  p} \times \boldsymbol q
- \alpha \frac{\boldsymbol q}{|\boldsymbol q|} = \boldsymbol {\dot q}
\times \boldsymbol {q} - \alpha \frac{\boldsymbol q}{|\boldsymbol q|}.
\end{equation}
To simplify
the reasoning, we put~$R=1$ and write the Lagrangian in the spherical
coordinates~\eqref{e12-29}
\begin{equation}
\label{e12-30}
L = \frac{1}{2}\left(\dot \theta^2 +
\sin^2 \theta \dot \varphi^2 + \sin^2 \theta \sin^2 \varphi \dot \psi^2\right) +
\alpha \cos \varphi \dot{\psi} - U(\theta).
\end{equation}

Suppose that the angular momentum vector \eqref{e12-28} is aligned with
the axis~$q_1^{}$ in the space~$q_1^{}$, $q_2^{}$, $q_3^{}$; then
\begin{equation}
\label{e12-31} M_2^{} = M_3^{} = 0,\quad \dot \varphi = 0, \quad \sin^2 \theta \dot
\psi = \frac{\alpha}{\cos \varphi_0^{}} = const.
\end{equation}

The latter relation is a generalization of Kepler's second law. Consider
the invariant surface in~$S^3$ given by~$\Bigl(\boldsymbol
M,\,\frac{\boldsymbol q}{|q|}\Bigr)= const$. On this surface, choose a
point, which is~$2\theta$ away from the particle. Then, the great-circle
arc, joining the origin of coordinates with the chosen point sweeps equal
areas in equal time intervals. (Indeed, the time rate of change of the
area is
$\frac{dS}{dt}=\biggl(\int_0^{2\theta}\sin\sigma\,d\sigma\biggr)
\frac{d\psi}{dt}=2\sin^2\theta\dot\psi.)
$

Taking into account the integral of energy~$E = h$ and \eqref{e12-31}, we
obtain
\begin{equation}
\label{e12-32} \dot \theta = \sqrt{2(h - U_c^{}(\theta))},\quad U_c^{}(\theta) =
U(\theta) + \frac{1}{2} \frac{\alpha^2 \theta n^2 \varphi_0^{}}{\sin^2 \theta}
\end{equation}
and the
explicit expression for the trajectory in terms of quadratures:
\begin{equation}
\label{e12-33}
\frac{\alpha\,d\theta}{\sin^2 \theta \sqrt{2(\tilde{h} - \tilde{U})} -
\frac{\alpha^2 \sin^2 \varphi_0^{}}{\sin^2 \theta}} = d\psi,
\end{equation}
 where $\tilde{h} = h
\cos^2\varphi^2_0$, $\tilde{U} = U \cos^2\varphi_0^{}$.

When $\gamma = 0$, an explicit expression in terms of quadratures for the
analog of the Poincar\'e problem is obtained from~\eqref{e12-33}, and the
trajectories are geodesics on the invariant cone defined by~$(\boldsymbol
M,\,\boldsymbol q/|\boldsymbol q|) = const$. As it was noted above, in the
case of the Appell problem's analog the trajectories are conic sections
(i.\,e. ellipses, hyperbolas, parabolas) on the plane development of the
cone. Generally, these are not closed on the cone, but there is a
possibility to ``adjust'' the potential \eqref{e12-12} so that the
trajectories will be always closed in the presence of a monopole. Due to
this ``tuning'', the Euclidean~MICZ-model can be generalized to the
spherical case, for which the Laplace--Runge--Lenz integral exists.

\section{Generalized MICZ-model}\label{sec5}

Consider a motion in the field of a monopole and in the field with the
potential
\begin{equation}
\label{e12-34}
U(\theta) = - \gamma\cot \theta +
\frac{1}{2}\frac{\mu}{\sin^2 \theta}, \quad \gamma, \mu = const.
\end{equation}
The trajectory in this case is given by~\eqref{e12-33}. If
\begin{equation}
\label{e12-35} \mu = \alpha^2,
\end{equation}
the trajectory is
\begin{equation}
\label{e12-36}
 \frac{\alpha d\theta}{\sin^2 \theta
\sqrt{2\tilde{h}+2\tilde{\gamma} \cot \theta - \frac{\alpha^2}{\sin^2 \theta}}} = d\psi.
\end{equation}

The trajectory \eqref{e12-36} is closed, and the gnomonic projection gives
us the conic section
\begin{equation}
\label{e12-37}
 \tan \theta  = \frac{p}{1 + e \cos(\psi -\psi_0)}
\end{equation}
with the following focal parameter and eccentricity:
$$
p=\frac{\alpha^2}{\gamma\cos^2\varphi_0},\quad
e=\sqrt{1 + \frac{2\alpha^2}{\gamma^2 \cos^2
\varphi_0^{}}\left(h - \frac{\alpha^2}{2 \cos^2 \varphi_0^{}}\right)}.
$$
The expression in terms of quadratures for $\dot \theta$ \eqref{e12-32} is
\begin{equation}
\label{e12-38}
\begin{split}
\dot \theta^2 = 2h + 2\gamma \cot \theta - \frac{\alpha^2}{\cos^2 \varphi_0^{} \cdot \sin^2
\theta}=\\
= 2h + 2 \gamma\cot \theta - \frac{c^2}{\sin^2 \theta} = f (\theta,\,c,\,h),
\end{split}
\end{equation}
where $c = {\alpha^2}/{\cos^2 \varphi_0^{}}$. Let us plot the bifurcational
diagram of the problem's solutions on the parameter plane~$(c^2,\,h)$. To
this end, recall that at critical points~$(c_\ast^{},\,h_\ast^{})$ on a
bifurcation curve, $f(\theta_0^{},\,c_\ast^{},\,h_\ast^{}) =
f'_0(\theta_0^{},\,c_\ast^{},\,h_\ast^{}) = 0$. As a result, we have two
curves (Fig. 1):
$$
{\rm I.}\;2h=c^2-\frac{\gamma^2}{c^2};\quad
{\rm II.}\;c^2=0.
$$

Besides, since~$c{\,=\,}\alpha^2/ \cos^2 \varphi_0^{}$, the
inequality~$c^2{\,>\,}\alpha^4$ also holds. Thus, in the plane defined by the
constants~$h,\,c^2$ (see Fig.~1), the domain of allowable values~$h,c^2$
lies above the line~$c^2 =\alpha^4$ and below the hyperbola defined by~I. For
a point of this plane above~$h = \frac{1}{2}c^2$, the particle moves only
in the upper half-plane of the sphere. Otherwise, the particle can also
move into the other half-plane.

It is easy to formulate an analog of Kepler's third law, coinciding with
the traditional law for a curved space~\cite{b12-13}. Indeed,
since~$\sin^2 \theta \dot \psi = c$, we have
$$ dt =
\frac{\sin^2 \theta d \psi}{c}. $$
 Therefore,
 \begin{equation}
 \begin{split}
\label{e12-39}
T = \frac{1}{c} \int_0^{2\pi} \sin^2 \theta (\psi) d \psi =
\frac{p^2}{c}\int_{-\pi}^{\pi} \frac{d\psi}{p^2 + (1 + e\cos \psi)^2} =\\
= \frac{\pi}{\sqrt{\gamma}}\sqrt{\frac{h}{\gamma} + \sqrt{1+
\frac{h^2}{\gamma^2}}}\cdot \left(1+\frac{h^2}{\gamma^2}\right)^{-1/2}.
\end{split}
\end{equation}
This dependence of the orbital period on energy can be easily transformed
into the dependence on (angular) length of the semi-major axis
\begin{equation}
\label{e12-40}
T = \frac{\pi}{\sqrt{\gamma}}\sqrt{-\tan a + \sqrt{1+ \tan^2
a}}(1 + \tan^2 a)^{-1/2},
\end{equation} where $\tan a = - \frac{\gamma}{h}$.

As in the Kepler problem (on~$\mathbb{R}$ and~$S^2$), closedness of the
trajectories is closely connected with some hidden symmetry of the
problem, i.\,e., with existence of a vector integral of the Laplace--Runge--Lenz type.

For the system~\eqref{e12-34}, \eqref{e12-35} this vector can be written
as
\begin{equation}
\label{e12-41}
\boldsymbol A = q_0^{} \tilde{\boldsymbol p} \times \boldsymbol M + \gamma R^2
\frac{\boldsymbol q}{|\boldsymbol q|}. \end{equation}

The Poisson brackets for the components of~$\boldsymbol A$ and the components of the
integral of angular momentum \eqref{e12-28} are:
\begin{equation}
\label{e12-42}
\begin{gathered}
\{M_i^{},\,M_j^{}\} = -\varepsilon_{ijk}^{} M_k^{},\quad \{M_i^{},\,R_j^{}\} = -
\varepsilon_{ijk}^{} R_k^{},\\
\{R_i^{},\,R_j^{}\} = 2\varepsilon_{ijk}^{} \left(R^2 h - \boldsymbol M^2 +
\frac{1}{2}\alpha^2\right)M_k^{}.
\end{gathered}
\end{equation}

As before, we can specify the conditions (in terms of the curvature of the
space and the value of the integral of energy), under which the symplectic
leaf~\eqref{e12-20} is compact.

\section{Trajectory isomorphism for central potential systems
on~$S^2$ and~$\mathbb{R}^2$}

For $U = U(r)$, the equations \eqref{e12-1} define a central potential
system on~$\mathbb{R}^3$. If~$U=U(\theta)$ in the
Lagrangian~\eqref{e12-13}, then we have a central potential system
on~$S^3$. These systems, respectively, have flat~($\mathbb{R}^2$) and
spherical~($S^2$) invariant manifolds. These two-dimensional systems can
be shown to be related, using the central (gnomonic) projection (from the
center of the sphere tangent to the plane at the attracting center) and
some suitable change of time.

Following Serret and Appell~\cite{b12-17,b12-18}, consider a system
in~$\mathbb{R}^2$ with the following equations of motion (in polar
coordinates):
\begin{equation}
\label{a1}
\frac{d}{dt}\left(\frac{\partial{T_p}}{\partial{\dot \rho}}\right) =
\mathrm{R}; \quad \frac{d}{dt}\left(\frac{\partial{T_p}}{\partial{\dot \varphi}}\right) = \Phi;
\end{equation}
where $T_p^{}$ is the kinetic energy of a point in the plane,
\begin{equation}
\label{a2}
T_p^{} = \frac{1}{2}\left(\dot \rho^2 + \rho^2 \dot \varphi^2\right),
\end{equation}
while $\mathrm{R}$, $\Phi$ stand for certain generalized forces (generally,
non-potential).

%\wfig<bb=0 0 56.4mm 27.3mm>{01_02.pcx}[The gnomonic projection]

Let us perform the transformation of coordinates (the gnomonic projection
from Fig.~2), forces and time:
\begin{equation}
\label{a3}
\begin{gathered}
\rho = \tan \theta,\quad \varphi = \psi,\quad dt = \cos^{-2}\theta d\tau\\
R = \cos^2 \theta\, \Theta,\quad \Phi = \cos^2 \theta\, \Psi.
\end{gathered}
\end{equation}

This results in a system on~$S^2$:
\begin{equation}
\label{a4}
\frac{d}{d\tau}\left(\frac{\partial{T_s}}{\partial\theta'}\right) =
\Theta,\quad
\frac{d}{d\tau}\left(\frac{\partial{T_s}}{\partial\psi'}\right) = \Psi,
\end{equation}
 where $\theta' =
\frac{d T_s}{d\tau}$, $\psi' = \frac{d\psi}{d\tau}$, while $T_s^{}$ is the
kinetic energy of a point on the sphere,
\begin{equation}
\label{a5} T_s^{} =
\frac{1}{2}\left(\theta' + \sin^2 \theta \psi'^2\right).
\end{equation} It is easy to see
that

{\bf Statement.} {\it
There exists a trajectory isomorphism between the Lagrangian system in
$\mathbb{R}^2$, with central potential}
$$
 L = \frac{1}{2}(\dot \rho^2 +
\rho^2 \dot \psi^2) + U(\rho), $$
{\it and the Lagrangian system on~$S^2$, with
central potential of the form}
$$ L = \frac{1}{2}(\dot \theta^2 + \sin^2
\theta \dot \psi^2) + U(\tan \theta). $$

To prove that, it is sufficient to put in~(\ref{a1})--(\ref{a5})
$$
\Phi = \Psi = 0$$

$$\mathrm{R} = - \frac{\partial{U}}{{\partial\rho}},\quad \Theta = -\frac{\partial{U}}{{\partial\theta}} =
-\frac{\partial{U}}{{\partial\rho}} \cdot
\frac{\partial{\rho}}{{\partial\theta}} = \frac{\mathrm{R}}{\cos^2 \theta}.$$

These transformations easily bring the plane Kepler problem to its analog
on sphere.

Note that under the transformation \eqref{a3} a potential field of forces
can be transformed into a non-potential field, and vice versa.

If we consider the inverse transformation~\eqref{a3} as a transformation
from sphere to plane, we have to adopt negative values of~$\rho$. In this
case,~$\rho$ is negative when the trajectory on the sphere crosses its
equator, or, in the plane ($\rho$, $\varphi$), when the trajectory jumps
from~${+\infty}$ to~${-\infty}$. If, instead of~$\rho = \tan \theta$, we
consider~$\xi = \cot \theta$, then the trajectory in the plane~($\xi$,
$\varphi$) is continuous.

It is also easy to show that the described isomorphism can be extended to
the generalized potential systems discussed in Sections~\ref{sec4},
\ref{sec5}. Note also that the transformation~(\ref{a3}), applied to the
Kepler problem on~$S^2$, can be used to generalize the Bohlin
(Levi--Civita) regularization. It can be shown that {\it on a fixed energy
level, the Kepler problem is reduced to the harmonic oscillator
problem\/}.

Another interesting property of the Kepler system in~$\mathbb{R}^2$ (due
to Hamilton) is that the velocity hodograph for a moving point is a circle
with a displaced center. A similar vector can be specified for the Kepler
problem on~$S^2$:
$$\boldsymbol \pi=\frac{\boldsymbol {\dot x}}{1+\boldsymbol {\dot{x}^2}/R^2},$$ where
$\boldsymbol x=(\rho\cos\varphi,\rho\sin\varphi)$ is the radius-vector of a point under
the gnomonic projection \eqref{a3}. The~$\boldsymbol \pi$ hodograph can be easily
shown to be a circle with a displaced center.

This work was supported from the program ``State Support for Leading
Scientific Schools'' (grant~NSh--36.2003.1); additional support was given
by the Russian Foundation for Basic Research (grant~04-05-64367) and the
CRDF (grant~RU-M1-2583-MO-04).

The authors thank Bruno Cordani for a copy of his book~\cite{b12-19}.

\begin{figure}[ht!]
$$
\includegraphics{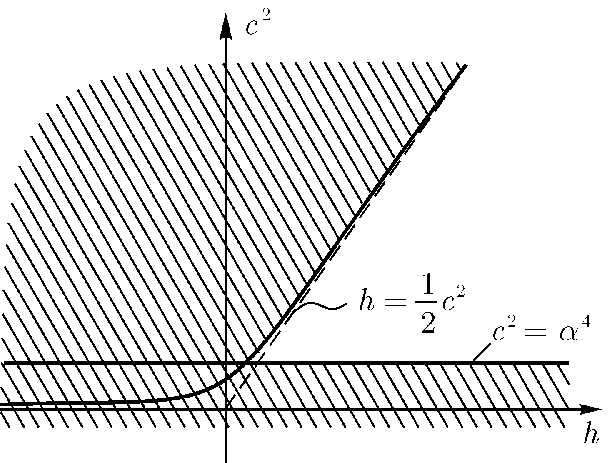}
$$
\caption{}
\end{figure}

\begin{figure}[ht!]
$$
\includegraphics{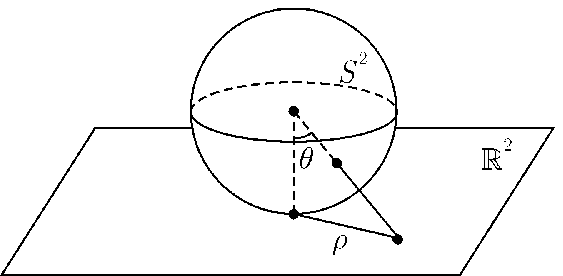}
$$
\caption{The gnomonic projection}
\end{figure}


\begin{thebibliography}{99}

\bibitem{b12-7}
Appell P. {\em Annales scientifical da Academia Polutechtnica}, do Porto,
V.\,IV, 1909, see also Appell P. {\em Traut\'e de m\'ecanique
rationnelle}, Paris, Gauthier-Villars, 1909.

\bibitem{b12-18}
Appell P. {\em Sur les lois de forces centrales faisant d\'{e}crire \`{a}
leur point d'application une conique quelles que soient les conditions
initiales}. American Journal of Mathematics. 1891. Vol. 13. p. 153-158.
%(See Ref.~\ref{ch6} in this book.)

\bibitem{b12-8}
Bates L. {\em Symmetry preserving deformations of the Kepler problem}.
Rep. Math. Phys., 1988, Vol. 26, p. 413-428.

\bibitem{b12-21}
Borisov\,A.\,V., Mamaev\,I.\,S. {\em Poisson structures and Lie algebras
in Hamiltonian mechanics.} Izhevsk: RCD, 1999 (in Russian).

\bibitem{b12-14}
Chernikov N.\,A. {\em The Kepler problem in the Lobachevsky space and its
solution}. Acta Phys. Polonica, 1992, Vol. 23, p. 115-119.
%(See Ref.~\ref{ch23} in this book.)

\bibitem{b12-20}
Cherno\"{\i}van\,V.\,A., Mamaev\,I.\,S. {\em The restricted two-body
problem and the Kepler problem in the constant curvature spaces.} Reg. \&
Chaot. Dyn., 1999, Vol.\,4, No. 2, p.\,112--124.

\bibitem{b12-19}
Cordani B. {\em The Kepler Problem}, Birkh\"{a}ser, 2003, 440~p.

\bibitem{b12-4}
Feher L.\,G. {\em Dynamical $O(4)$ symmetry in the asymptotic field of a
Prasad-Sommerfield monopole}. J.~Phys.~A, 1941, Vol. 19, p. 83-89.

\bibitem{b12-1}
Fok V.\,A. {\it Hydrogen Atom and Non-Euclidean geometry}. Izv. AN SSSR,
OMEN, 1935, Vol. 2, p. 169-179 (in Russian).

\bibitem{b12-12}
Granovsky Ya.\,I., Zhedanov A.\,S., Lutsenko I.\,M. {\em Quadratic
Algebras and Dynamics in Curved Space\/}. I. {\it Oscillator}\/. II. {\it
The Kepler Problem}. Theor. and math. phys., 1992, Vol. 91, Nos. 2; 3,
pp.~207-216; 396-410 (in Russian). %(See Ref.~\ref{ch15} in this book.)

\bibitem{b12-2}
Gy\"{o}rgyi G. {\em Kepler's equations, Fock variables, Bacry's generators and
Dirac brackets.} Nuovo Cimento, 1968, Vol. 53A, p. 717-735.

\bibitem{b12-11}
Higgs P.\,W. {\em Dynamical symmetries in a spherical geometry}. I.
J.~Phys.~A., 1979, Vol. 12, No. 3, p. 309-323. %(See Ref.~\ref{ch9} in this book.)

\bibitem{b12-13}
Killing W. {\em Die Mechanik in den Nicht-Euklidischen Raumformen.}
J.~Reine Angew. Math, 1885, Vol. 98, p. 1-48.
%(See Ref.~\ref{ch3} in this book.)

\bibitem{b12-9}
Kozlov V.\,V. {\em On Dynamics in Constant Curvature Spaces}. Vestnik
MGU, ser. math. mech., 1994, No. 2, p. 28-35 (in Russian). %(See Ref.~\ref{ch10} in this book.)

\bibitem{b12-16}
Kozlov V.\,V., Harin A.\,O. {\em Kepler's problem in constant curvature
spaces}. Celestial Mech. and Dynamical Astronomy, 1992, Vol. 54, p.
393-399. %(See Ref.~\ref{ch11} in this book.)

\bibitem{b12-15}
Leach P.\,G.\,L., Fleassas G.\,P. {\em Generalizations of the Laplace-Runge-Lenz Vector.} Journal of Nonlinear Mathematical Physics, 2003,
Vol. 10, No. 3, p.~340-423.

\bibitem{b12-10}
Liebmann Í. {\em \"{U}ber die Zantalbewegung in der nichteuklidiche
Geometrie}. Leipzig Ber, 1903, Vol. 55, p. 146-153.
%(See Ref.~\ref{ch4} in this book.)

\bibitem{b12-6}
McIntosh Í., Cisneros A. {\em Degeneracy in the presence of a magnetic
monopole}. J.~Math. Phys., 1970, Vol. 11, p. 896-916.

\bibitem{b12-3}
Moser J. {\em Regularization of Kepler's problem and the averaging method on a
manifold.} Comm. Pure Appl. Math., 1970, Vol. 23, p. 609-636.

\bibitem{b12-17}
Serret P. {\em Th\'{e}orie nouvelle g\'{e}om\'{e}trique et m\'{e}canique des
lignes a double courbure}. Paris, Librave de Mallet-Bachelier, 1860.
%(See Ref.~\ref{ch1} in this book.)

\bibitem{b12-5}
Zwanziger D. {\em Exactly soluble nonrelativistic model of particles with
both electric and magnetic charges}. Phys. Rev., 1968, Vol. 176, p. 1480-1488.

\end{thebibliography}
\end{document}